# Photon number resolving detection at telecom wavelength with a charge integration photon detector

**Mikio Fujiwara and Masahide Sasaki**

Address

Basic and Advanced Research Department, National Institute of Information and Communications Technology, 4-2-1 Nukui-kita, Koganei, Tokyo 184-8795, Japan

We demonstrate multiphoton discrimination at telecom wavelength with the readout frequency of 40 Hz by charge integration photon detector (CIPD). The CIPD consists of an InGaAs pin photodiode and a GaAs junction field effect transistor as a pre-amplifier in a charge integration circuit, which is cooled to 4.2 K to reduce thermal noise. The quantum efficiency of the CIPD (detector itself) is 80% for 1530 nm light, and the readout noise is measured as 0.26 electrons at 40 Hz. We can construct Poisson distributions of photo-carriers with distinguished peaks at each photo-carrier number corresponding to the signal to noise ratio of about 3.

A photon number resolving detector (PNRD) that enables to count the number of photons in a pulse precisely is an important tool to develop universal photonic quantum gates.[1-3] Even small scale photonic quantum computers implemented in this way will be used to construct quantum decoders that can overcome the conventional capacity limit of a communications channel.[4] PNRDs in telecom fiber wavelengths are important in this regard. In addition, it can also improve the security of quantum key distribution systems.[5,6] Obviously, PNRDs can bring many benefits for research fields such as bioinstrumentation, environmental remote sensing, and astronomy. Concretely speaking, in observation of active Oxygen in animal and vegetable dietary fibers[7], and differential absorption lidar measurements of $CO_2$ and $O_2$ in the Air[8], and deep survey of universe[9], ultra sensitive near infrared detectors will bring new and accurate data by detecting faint light (<1000 photons/s). PNRD technology, however, is not acceptable for those wide applications. An InGaAs APD single photon detector with the gated Geiger mode employed in quantum cryptography in telecom wavelengths cannot distinguish between one and more than one photon and loses the information of photon number, that prevents us to directly detect non-classical states of light. Moreover, its quantum efficiency is about 10~20%, and the counting rate is also limited to about 1 MHz due to the after pulse phenomenon.[10] A superconducting transition-edge sensor (TES)[11,12], on the other hand, can be a hopeful candidate as a PNRD at telecom wavelength. Its quantum efficiency has recently been improved to 80% with resonant cavities.[12] However, the tungsten TES must be cooled to below 100 mK, and special readout technique such as a SQUID array amplifier is needed. Moreover, the TES has sensitivity for a wide range of wavelength. While this character is desirable for spectroscopy, it also means that it suffers from background radiation. So very tight radiation shield is necessary. These circumstances will hinder diffusion of the TES.

As another candidate for a PNRD at telecom wavelengths, we proposed a charge integration photon detector (CIPD)[13] constructed with an InGaAs PIN photodiode (Kyosemi Corp., Japan) and a GaAs JFET (SONY Ltd., Japan) as a pre-amplifier of a charge integration circuit. In our previous work[13], we demonstrated the photon number discrimination capability with the signal to noise ratio (*S/N*) of more than 2. The quantum efficiency was 80%, but the repetition rate was still 1 Hz. Although the readout speed remains slow at present, the CIPD can be fabricated with commercially available components, and no special technology is needed except cooling down at 4.2 K. This range of temperature can be attained with an electrical intercooler these days, and even a 77 K radiation shield is enough for darkening of the CIPD. In this letter, we report an examination for operating high readout frequency (40 Hz) of the CIPD and describe the accuracy of detection.

Figure 1(a) shows the conceptual view of our CIPD. The InGaAs PIN photodiode used has high quantum efficiency (>80%) and good linearity for incident photons without multiplication noise. Photo-carriers that are accumulated at the gate electrode of the GaAs JFET increase the source voltage due to the source follower connection. The source electrode is connected to an amplifier (voltage follower and Stanford Research System: model SR560) at room temperature. The framed area in Fig. 1(a) is cooled to 4.2 K to reduce noise and leakage current. Photo-carriers in the gate electrode are reset through a mechanical probe <u>driven by a solenoid</u>, when the output voltage increases to the region in which the source follower gain decreases. The *S/N* in this CIPD is given as follows:

$$S/N = G_M \cdot NQ \Big/ \left( C_{imput} \cdot V_{noise,CDS} \right) \qquad (1)$$

where $G_M$ is the source follower gain, $C_{input}$ is the input capacitance of the CIPD, $Q$ is the elemental electric charge, $N$ is the number of photo-carriers, and $V_{noise,CDS}$ is the channel noise of the GaAs JFET at correlation double sampling.[13] This equation clearly specifies that reducing input capacitance and system noise are essential to improve resolution and speed up readout frequency. In our previous work, junction size of the GaAs JFET was 5 μm width and 50 μm length. Such a long gate length was chosen for the purpose of reducing $1/f$ noise and random telegraph signal (RTS) noise. We, however, found that even for a shorter gate length such as 10 μm, and low frequency noise such as $1/f$ and/or RTS noise in the GaAs JFET can be suppressed down to ~500 nV/Hz$^{1/2}$ at 1 Hz by using a noise reduction method with thermal process, referred to as the thermal cure.[14] Actually it was argued that these noises could be mainly attributed to the band bending and deep-level traps near the drain electrode and it was found that there was an optimal operation condition where the noises do not increase much for a smaller size.[15] The input capacitance of the GaAs JFET decreases from 0.041 pF to 0.037 pF. Moreover, the transconductance increases by shortening the gate length, and as a result $G_M$ is improved from 0.85 to almost 1. The capacitance of a new InGaAs PIN photodiode whose diameter is 30 μm decreases from 0.026 pF to 0.017 pF at 4.2 K by using a high dopant concentration type, although this result contradicts the commonly observed fact at room temperature. Totally, the input capacitance can be reduced to 0.054 pF, and one photo-carrier generates 3 μV as the output voltage step.

The noise spectra of CIPDs which were measured using an Advantest FFT analyzer (model 9211) are shown in Fig. 1(b). Solid line is for the present CIPD, while dashed line is for the previous one, respectively. The noise level of the new one is smaller than 500 nV/Hz$^{1/2}$ at 1 Hz and dips from the previous one on average up to 50 Hz. The direct measurement of the standard deviation of the output voltage in the dark condition at 40 Hz results in 0.26 electrons shown in Fig. 1(c). This result implies that we can discriminate 0 or 1 electron with *S/N* of ~4. Leakage current of the CIPD is 500~1000 electron/hour.

The cutoff wavelength of the InGaAs PIN photodiode at 4.2 K becomes short as 1540 nm.[13] Therefore, for testing photon counting, heavily attenuated coherent light pulses at 1530 nm are used. Light from the CW laser is shaped into a pulse with a 2.5 msec width and repetition frequency of 40 Hz by a LN EO modulator. Photons are coupled to the detector through a single mode fiber with a focuser, and coupling efficiency is up to 80%.

Figure 2 shows the histograms of the measured photo-carriers distributions at a sampling of ~700 events with resolution of 0.1 electrons. This resolution is limited by uncertainty in the measurement of $C_{input}$ or $G_M$, and this digit includes error. Lines in the Fig. 2 are fitting curves given as follows,

$$N(x) = \frac{1}{\sqrt{2\pi}\sigma} \sum_{l}^{max} P_n(l) \exp\{-(x-l)^2/2\sigma^2\} \qquad (2)$$

where $P_n$ is Poisson distribution with the average photo-carriers of *n*, *max* is a cutoff number of photo-carriers (in this case 20), and $\sigma$ is the standard deviation reflecting the noise characteristics of the readout circuit. In Fig. 2, $\sigma$ is evaluated as 0.3~0.33. Multipeak structures can be recognized up to 4 photo-carrier numbers, well corresponding to the peak positions of the curves of Eq. (2). This result also corroborates that our estimation of the input capacitance was done precisely. Overlaps of the peaks correspond to the error of photon number discrimination with the *S/N*~3.

In Fig. 3, we show histograms of the photo-carrier number distribution with resolution of 1 electron for six kinds of pulse intensities. The histograms are in good agreement with systematic changes of the theoretical curves of Poisson distributions up to the number of 10. The quantum efficiency is estimated as 80±5% by back calculation.

We examine the charge integration photon detector with an InGaAs PIN photodiode and a cryogenic GaAs JFET as a photon number resolving detector for a telecom wavelength, and demonstrate that the CIPD can measure photo-carrier number with the resolution of the $S/N$~3 and the quantum efficiency of 80% at 40 Hz. This CIPD has potential of precise spectroscopy for many research fields. In addition, our readout circuit based on the charge integration amplifier can also be applied not only to the readout of photo-carriers but also to accurate determination of minimal charges and the capacitance at cryogenic temperature, required, for example, in device physics.

References:

1. N. Knill, R. Laflamme, and G.J. Milburn, Nature **409**, 46 (2001).

2. D. Gottesman, A. Kitaev, and J. Preskill, Phys. Rev. A**65**, 012310 (2001).

3. S. D. Barlett and B. C. Sanders, Phys. Rev. A**65**, 042304 (2002).

4. M. Fujiwara, M. Takeoka, J. Mizuno, and M. Sasaki, Phys. Rev. Lett. **90**, 167906 (2003).

5. G. Ribordy, J. D. Gautier, N. Gisin, O. Guinnard, and H. Zbinden, J. Modern Opt. **47**, 517 (2000).

6. G. Ribordy, J. Brendel, J. D. Gautier, N. Gisin, and H. Zbinden, Phys. Rev. A**63**, 012309 (2000).

7. N. Suzuki, A. Fujimura, T. Nagai, I. Muzumoto, T. Itami, H. Hatate, T. Nozawa, N. Kato, T. Nomoto, and B. Yoda, BioFactors, **21**, 329 (2004).

8. P. F. Ambrico, A. Amodeo, P. D. Girolamo, and N. Spinelli, Appl. Opt., **39**, 6847 (2000).

9. G. G. Fazio *et. al*., Astrophys. J. **154**, 10 (2004).

10. A. Tomita and K. Nakamura, Opt. Lett. **27**, 1827 (2002).

11. A. J. Miller, S. W. Nam, J. M. Martinis, and A. V. Sergienko, Appl. Phys. Lett., **83**, 791 (2003).

12. D. Rosenberg, A. E. Lita, A. J. Miller, S. Nam, and R. E. Schwall, IEEE Trans. Appl. Supercond., **15**, 575 (2005).

13. M. Fujiwara and M. Sasaki, Appl. Phys. Lett., **86**, 111119 (2005).

14. M. Fujiwara, M. Sasaki, and M. Akiba, Appl. Phys. Lett., **80**, 1844 (2002).

15. M. Fujiwara and M. Sasaki, IEEE Trans. Elec. Dev. **51**, 2042 (2004).

Figure Caption

**FIG. 1.** (a) Conceptual cross sectional views of the charge integration photon detector (CIPD). Insert shows the picture of the cryogenic circuit. (b) Noise spectra of the CIPDs at 4.2 K. Solid line is for present CIPD's, while dashed line is for the previous ones. (c) Distribution of the dark counts at 40 Hz by direct measurement. The standard deviation is 0.26.

**FIG. 2.** Histograms of pulse heights for (a) average photo-carriers: 1.07 (b) average photo-carriers: 2.55, (c) average photo-carriers: 2.85.

**FIG. 3.** Photo-carrier number distribution: average photo-carriers is (a) 1.58, (b) 1.84, (c) 2.22, (d)3.07, (e) 4.01, (f) 10.18. Lines in figures are theoretical curve of Poisson's distributions including the noise characteristics of the readout circuit.

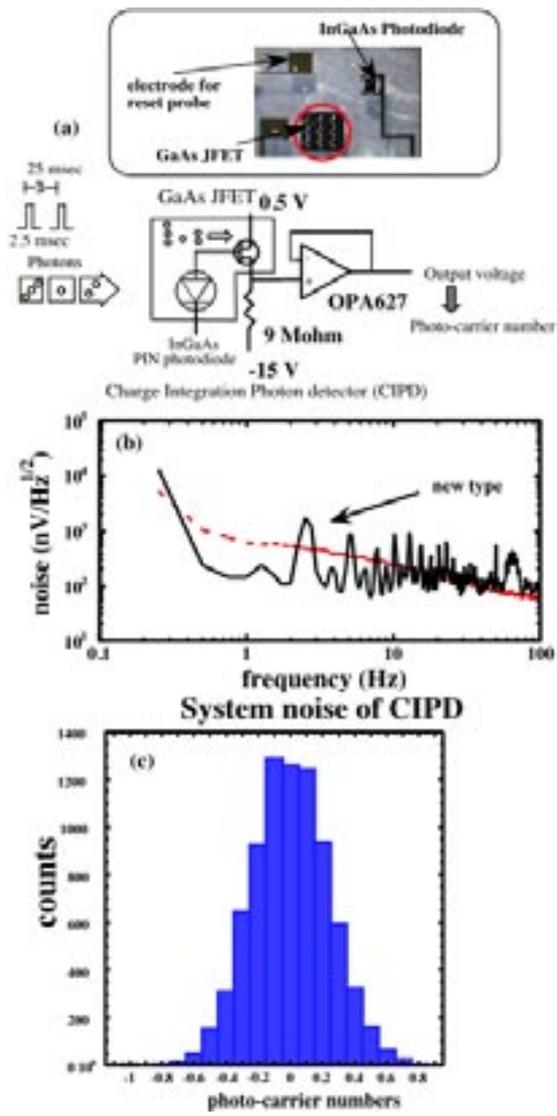

FIG. 1
9

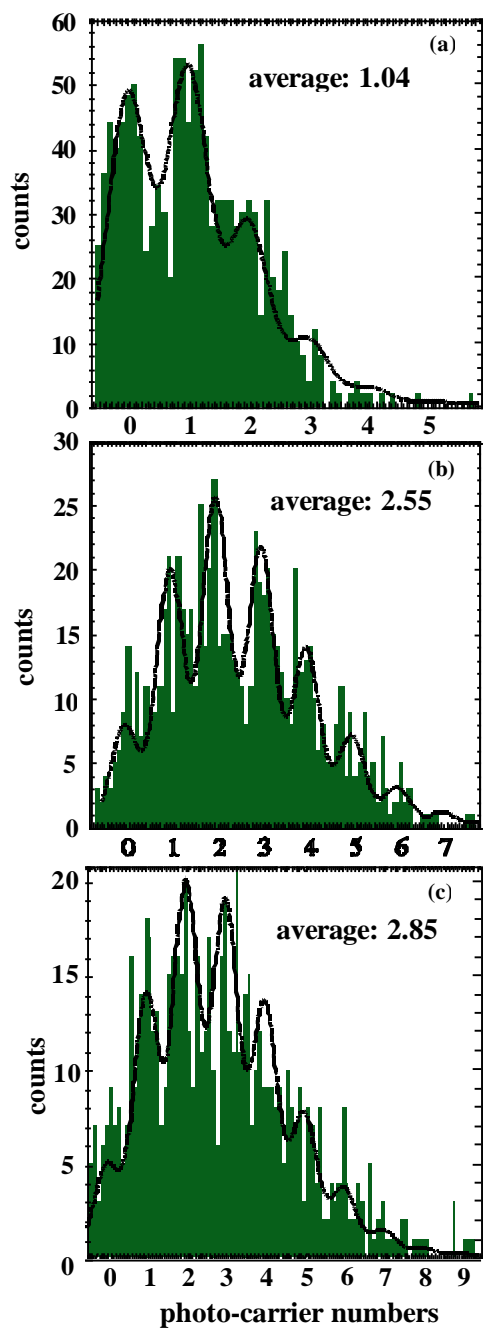

FIG. 2

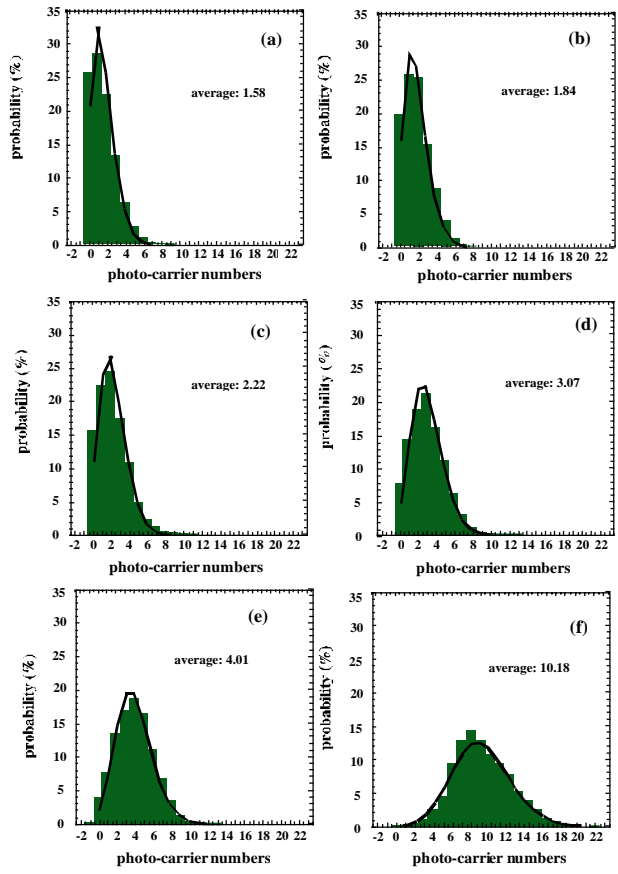

FIG. 3